\documentclass[10pt]{iopart}

\usepackage{graphicx}
\usepackage{latexsym} 

\begin{document}

\title{Sub-threshold $\phi$ and $\Xi^-$ production by high mass resonances with UrQMD}

\author{J. Steinheimer$^1$ and M. Bleicher$^{1,2}$}

\address{$^1$ Frankfurt Institute for Advanced Studies, Ruth-Moufang-Str. 1, 60438 Frankfurt am Main, Germany}
\address{$^2$Institut f\"ur Theoretische Physik, Goethe Universit\"at Frankfurt, Max-von-Laue-Strasse 1, D-60438 Frankfurt am Main, Germany}
\ead{steinheimer@fias.uni-frankfurt.de}
\vspace{10pt}

\begin{abstract}
We present a possible explanation for the deep sub-threshold, $\phi$ and $\Xi^-$ production yields measured with the HADES
experiment in Ar+KCl reactions at $E_{\mathrm{lab}}=1.76$ A GeV and present predictions for Au+Au reactions at $E_{\mathrm{lab}}=1.23$ A GeV. To explain the surprisingly high yields of $\phi$ and $\Xi^-$ hadrons we propose new decay channels for high mass baryon resonances. These new decay channels are constrained by elementary $\mathrm{p+p}\rightarrow \mathrm{p+p+}\phi$ cross sections, and $\Xi^-$ production in p+Nb. Based on the fits to the elementary reactions one obtains a satisfactorily description of $\phi$ and $\Xi^-$ production in deep sub-threshold Ar+KCl reactions as well as the observed nuclear transparency ratio in proton induced $\phi$ production in cold nuclear matter. The results implicate that no new medium effects are required to describe the rare strange particle production data in low energy nuclear collisions.
\end{abstract}

%
%
\submitto{\jpg}
%
%
\ioptwocol

\section{Introduction}
The study of strange hadrons in nuclear collisions has since long been considered to be a good probe for the properties of dense hadronic matter \cite{Koch:1986ud,Aichelin:1986ss}. As strangeness in nuclear collisions has to be newly produced (as $s+\overline{s}$ pair), and the strange quark mass is significantly larger than that of the light quarks, it is possible to study near and sub-threshold production of strange hadrons in nuclear collisions at energies where systems of large net baryon density are created \cite{Randrup:1980qd}. The properties of such hot and dense systems are in the focus of current and planned experimental programs at the GSI/FAIR \cite{Friman:2011zz}, NICA \cite{nicawhitepaper} and RHIC facilities. Because ab initio calculations of QCD (for example lattice QCD) are not applicable at such high baryon densities \cite{Allton:2002zi,Kaczmarek:2011zz,Borsanyi:2012cr,Borsanyi:2013hza,Gavai:2008zr} one has to rely on phenomenological models and comparisons to data to learn about the properties of the dense matter created in these experiments. For this purpose microscopic transport models are usually employed. Several previous works found that the production rates and properties of Kaons, for example are a promising probe to extract their medium interactions in low energy nuclear collisions \cite{Aichelin:1986ss,Shor:1989nz,Hartnack:1993bq,Fang:1994cm,Li:1994cu,Li:1994vy,Mosel:1992rb,Miskowiec:1994vj,Cassing:1996xx,Bratkovskaya:1997pj,Hartnack:2001zs,Hartnack:2005tr,Hartnack:2011cn}. The before mentioned transport models successfully describe Kaon production in low energy nuclear collisions by modifying the medium properties of strange hadrons. However an important difference to the UrQMD model, which we will use in our study, is that in UrQMD a significant larger number of high mass resonances is included.  

A rather phenomenological approach to strangeness production is the so called thermal model. This model is usually based on a (grand)-canonical fit to experimentally measured particle ratios or yields and gives surprisingly good results for hadron production at high energies, i.e. $\sqrt{s_{NN}}> 5$ GeV (see e.g. \cite{Becattini:2000jw,Andronic:2005yp}). At the low near and sub-threshold energies under investigation at the SIS18 accelerator however, it was found that the thermal model has significant problems describing the set of strange particle ratios measured with the HADES experiment \cite{Agakishiev:2015xsa,Agakishiev:2009ar,Agakishiev:2009rr,Kolomeitsev:2012ij}. This poses the question whether one has to take into account new and interesting physics to describe strange particle production at these low beam energies. 

In this work we will explore new mechanisms for the production of the $\phi$ meson and $\Xi$ baryon
in the microscopic transport model UrQMD. In the UrQMD model most of the particle production at lower beam energies 
$\sqrt{s_{NN}} \leq 4$ GeV takes place by resonance excitation and decay. Therefore we explore new decays 
of the sort $N^{*}\rightarrow N +\phi$ and $N^{*}\rightarrow \Xi +K + K$ for the most heavy resonances. Evidence for such channels has recently been put forward in \cite{Fabbietti:2015tpa}. As the experimental knowledge of these exotic decays of very heavy baryon resonances is scarce, one has to infer the decay probabilities indirectly. The probabilities for these decays can be fixed using near threshold production in elementary reactions. Once the production probabilities are fixed we use these to estimate multi-strange hadron production in sub-threshold nuclear collision in order to determine baseline production estimates consistent with elementary reactions, i.e. vacuum physics.

The paper is organized as follows. 
First we will discuss possible mechanisms of particle production in UrQMD, with an emphasis on $\phi$ and $\Xi^-$ hadrons. Then we will outline how we introduce a new production channel for these hadrons through the decay of massive $N^*$ states. We will show how the introduction of these new decays can be used to describe elementary production cross sections near their threshold energies and finally study the effect of these production channels on the deep sub-threshold production of multi-strange hadrons in nuclear collisions. 

\section{Particle production in UrQMD}
\begin{figure}[t]	
\includegraphics[width=0.5\textwidth]{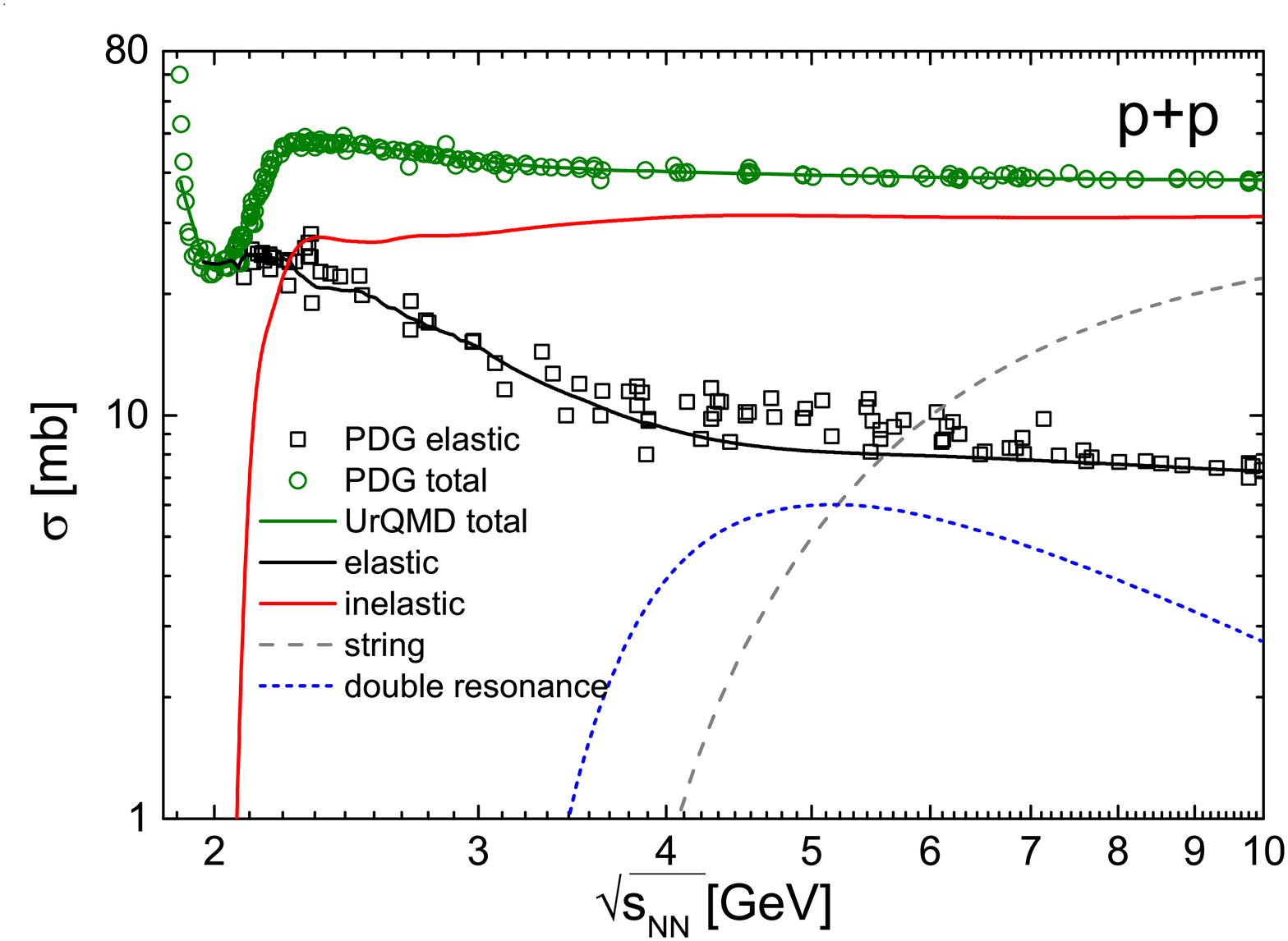}	
\caption{[Color online] Total (green line) and elastic (black line) proton+proton cross sections implemented in the UrQMD model compared with data from \cite{Agashe:2014kda} (symbols). We also show the total inelastic (red line) cross section. The parts of the inelastic cross section which correspond to the string excitation (grey dashed line) and the double Resonance $NN\rightarrow R^* R^*$ channel (blue dashed line) are also shown.}\label{f0}
\end{figure}

Hadron production in the UrQMD transport model \cite{Bass:1998ca,Bleicher:1999xi} proceeds through different channels: The excitation and de-excitation (decay) of hadronic resonances, the excitation and de-excitation of a string and the annihilation of a particle with its anti-particle. The probabilities of the different processes are governed by their reaction cross sections. These cross sections serve as input for the model and are, whenever possible, taken from experimental measurements of elementary (binary) collisions. For example the total and inelastic cross section of binary proton+proton collisions has been measured in many experiments over a wide range of beam energies \cite{Agashe:2014kda} (see green circles and black squares in figure \ref{f0}). The corresponding green and black lines in fig. \ref{f0} serve as input for model. The difference between the total and elastic cross section 
therefore should correspond to the inelastic cross section. Here again, many different reactions are possible. In UrQMD the inelastic part of the nucleon+nucleon cross section (up to a certain energy) is described by resonance production channels. The possible channels of resonance excitations are divided into several classes:
\begin{enumerate}
\item $NN\rightarrow N\Delta_{1232}$
\item $NN\rightarrow NN^{*}$
\item $NN\rightarrow N\Delta^{*}$
\item $NN\rightarrow \Delta_{1232}\Delta_{1232}$
\item $NN\rightarrow \Delta_{1232}N^{*}$
\item $NN\rightarrow \Delta_{1232}\Delta^{*}$
\item $NN\rightarrow R^* R^*$
\end{enumerate}    

Here $R^*$ could be any excited $N^*$ or $\Delta^*$ state.
Since a large part of the channels are not known, or only measured within a limited energy interval one uses an effective parametrization of the different cross sections:

\begin{equation}
\sigma_{1,2\rightarrow3,4}(\sqrt{s}) \propto (2S_3+1)(2S_4 +1) \frac{\left\langle p_{3,4}\right\rangle}{\left\langle p_{1,2}\right\rangle}\frac{1}{s} \left|M(m_3,m_4)\right|^2
\end{equation}

where $\left\langle p_{i,j}\right\rangle$ is the average momentum of the in- and outgoing particles and $S_i$ are the spins of the outgoing particles. The matrix element $\left|M(m_3,m_4)\right|^2$ is usually not known for all reactions but can be parametrized to fit experimental measurements, if available. A detailed description on how the elementary cross sections of processes 1 to 6 can be parametrized as well as comparisons with data is shown in detail in \cite{Bass:1998ca,Bleicher:1999xi}. The process $NN\rightarrow R^* R^*$ has just recently been introduced into the model with:
\begin{equation}
	\left|M(m_3,m_4)\right|^2= \frac{A}{1+ (m_4-m_3)^2(m_4+m_3)^2}
\end{equation}
The parameter $A=0.05$ was chosen such that the double resonance excitation is consistent with the total and inelastic (total-elastic) part of the p+p cross section. The contribution of double resonance excitation to the p+p cross section is depicted as the blue short-dashed line in figure \ref{f0}.

However, at some center-of-mass energy resonance excitation becomes unable to describe the inelastic cross section measured in experiment. This increasing difference between the resonance channels and the inelastic cross section is filled with the string excitation channel, which is also shown in figure \ref{f0} as dashed grey line. One can see that the string channel starts to dominate particle production at beam energies above $7$ to $10$ GeV. Since we are interested in near and sub-threshold particle production the string channel is not relevant for most of the following results, whereas the resonance channels are essential for an understanding of the beam energies we will consider.

Another important channel for the description of strange particle production in nuclear collisions is the strangeness exchange reaction which can change the flavor content of a hadron. This includes reactions of the type $N+\overline{K}\leftrightarrow Y + \pi$ as well as $Y+\overline{K} \leftrightarrow \Xi + \pi$. Such reactions are included in the UrQMD transport approach, however it was shown that they are not sufficient to explain the large $\Xi/\Lambda$ ratio in Ar+KCl and p+Nb reactions measured with the HADES experiment (see \cite{Graef:2014mra} for a detailed discussion). In fact the large $\Xi/\Lambda$ ratio in near-threshold p+Nb reactions, where strangeness exchange reactions cannot contribute, indicates that $\Xi$ production occurs as a single step (direct production) process.\\

\section{On the possibility of $\phi$ and $\Xi$ production}  
  
When discussing sub-threshold production of $\phi$'s and $\Xi$'s in nuclear collisions one should 
note that there are two distinct mechanisms which allow for the production of hadrons with masses, higher
than what would be energetically forbidden in elementary reactions:
\begin{enumerate}
\item One is the fact that in a nucleus, 
the nucleons acquire a Fermi momentum due to their bound state. Because of the Fermi momenta, the 
actual energy of two colliding nucleons will not be exactly the beam energy but a smeared out energy distribution.
This allows for collisions of nucleons at energies higher than the actual beam energy.
\item
Furthermore energy can be accumulated due to secondary interactions of already excited states, produced earlier in the collision \cite{Spieles:1993jx,Zeeb:2003wv}.
\end{enumerate}
As an example for deep sub-threshold production of multi-strange hadrons we will first investigate the production probability of resonance states with sufficiently high mass to produce a $\phi$ or $\Xi$ in collisions of Ca+Ca (corresponding to the
Ar+KCl collisions studied at the HADES experiment). In such collisions, from the Fermi momenta alone, about two percent of all primary $N+N$ collisions will have an invariant mass large enough to produce a $\phi$ meson, while essentially none has sufficient energy to produce a $\Xi$. Even though a small fraction of initial collisions has sufficient energy to produce a $\phi$, none can be produced at this energy through a string excitation because in the string picture a $\phi$ can only be created together with a $K+\overline{K}$ pair, increasing the effective threshold for $\phi$ production by a string fragmentation by an additional 1 GeV.\\

\begin{figure}[t]	
\includegraphics[width=0.5\textwidth]{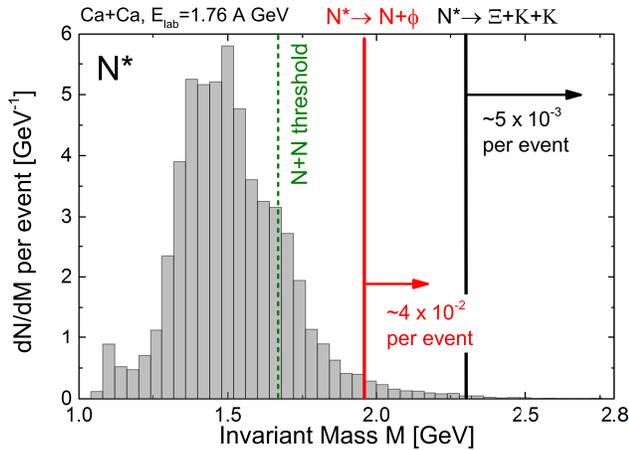}	
\caption{[Color online] Invariant mass distribution of $N^*$ resonances produced in collisions of Ca+Ca at a fixed target beam energy of $E_{\mathrm{lab}}= 1.76$ A GeV. We consider events with an impact parameter smaller than $b<5$ fm, in accordance with HADES experiment specifications. The vertical green dashed line indicates the maximum mass a $N^*$ can have in an elementary $N+N$ collision at the same beam energy. The vertical red line depicts the $\phi$ production threshold mass while the black line corresponds to the $\Xi+K+K$ threshold mass. 
}\label{f1}
\end{figure}		

Figure \ref{f1} shows the calculated invariant mass distributions of $N^*$ baryons 
produced in Ca+Ca collisions at a fixed target beam energy of $E_{lab}=1.76$ A GeV and a centrality range of $b<5$ fm. The vertical green dashed line indicates the maximal $N^*$ mass possible in elementary $NN$ reactions at the same beam energy.
It is clear from this figure that a substantial number of $N^*$ resonances with masses larger than the apparent threshold energy is produced in the nuclear collision. We also indicate, as vertical lines, the minimal mass a $N^*$ would need in order to decay into a final state with a $\phi$ (red line) or even $\Xi$ (black line). From this distribution we conclude that a moderate amount of 
excited states with sufficiently high mass are available that may produce $\phi$ mesons as well as $\Xi$ baryons. In the following we explore decays of the most massive $N^*$ resonances implemented in UrQMD, namely the $N^*\rightarrow N+ \phi$ and $N^*\rightarrow \Xi+ K +K$ channels, in order to describe the production of $\phi$ and $\Xi$ particles near and below their elementary threshold energies. In particular we will
use the $N^*(1990)$, $N^*(2080)$, $N^*(2190)$, $N^*(2220)$ and $N^*(2250)$ states included in the UrQMD model, as their decay channels are experimentally not well constrained and they have a sufficiently large mass.
One should note that, by introducing these new decay channels we also 
naturally, through detailed balance relations, introduce reactions of the kind $M+N \leftrightarrow N^* \leftrightarrow N+ \phi$, where $M$ could be any meson that couples to the $N^*$ (e.g. $\eta$, $\omega$, $\rho$ or $\pi$). Such channels have been discussed in \cite{Chung:1997mp} as possible sourced of $\phi$ mesons in nuclear collisions. In \cite{Chung:1997mp} however, the authors determined the relevant cross sections from theoretical models and found that such processes cannot fully account for the measured $\phi$ yield. Here we take a different approach and extract the cross sections directly from the comparison with experimental data, i.e. the resonance production in p+p reactions. This has the advantage that we do not have to rely on the validity of certain model assumptions, while on the other hand we can not predict the $\phi$ meson production in elementary reactions from fundamental calculations. 
   \begin{figure}[t]	
\includegraphics[width=0.5\textwidth]{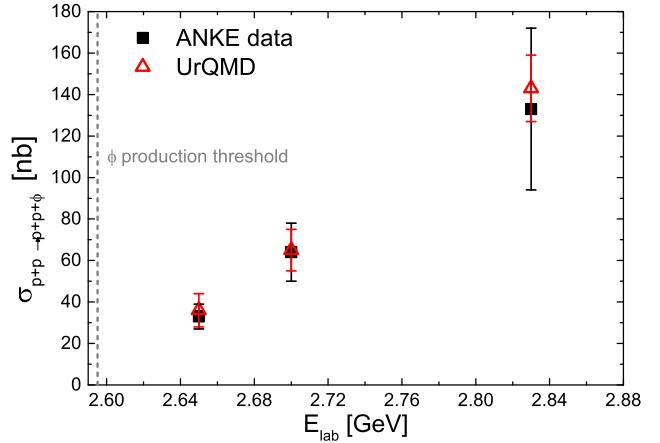}	
\caption{[Color online] Total production cross section of the $\phi$ meson in p+p reactions near the $\phi$ production threshold. We compare experimental data from \cite{Maeda:2007cy} with UrQMD results including the $\phi$ decay of the $N^*$ resonances.
}\label{f2}
\end{figure}		

\section{Elementary reactions}

As a next step we need to determine the probability that a heavy baryonic resonance state decays into the specific final states introduced above. Fortunately the ANKE experiment has recently published a set of data on the cross section of single $\phi$ production
in near threshold $p+p$ collisions \cite{Maeda:2007cy}. These cross sections are shown in figure \ref{f2} as black squares.
Using these data we find, that a branching fraction of $\frac{\Gamma_{N+\phi}}{\Gamma_{\mathrm{tot}}} = 0.2 \%$, for all the above mentioned $N^*$ resonances, provides a very good description (red triangles) of the measured $\phi$ production cross section. Note that we only fit one parameter, the branching fraction, to obtain a description of the data for all three measured beam energies.  
One should further note that the extracted branching fraction of $0.2 \%$ is roughly two orders of magnitude smaller than that of other channels like the $N^* \rightarrow N+\omega$ dacay and therefore compatible with the OZI supression factor experimentally extracted from the $\phi/\omega$ production ratio in $p+p$ collisions \cite{Sibirtsev:2005zc}.\\

\begin{figure}[t]	
\includegraphics[width=0.5\textwidth]{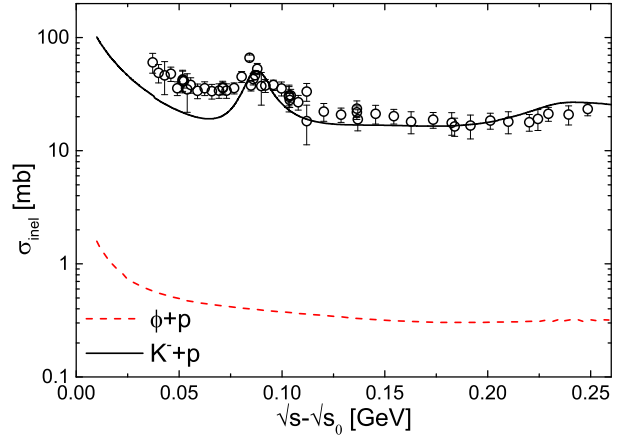}	
\caption{[Color online] Inelastic scattering cross sections for $K^-$ (black solid line) and $\phi$ (red dashed line) mesons with protons as function of the excess invariant mass. The $K^-$ cross section is compared to the difference of total and elastic cross sections taken from the PDG compilation (black symbols).   
}\label{f9}
\end{figure}

As we discussed above, the new decay channel, through the detailed balance relations, introduced an inelastic  $\phi+N \leftrightarrow N^*$ absorption channel for the $\phi$ meson. The cross section obtained from our fit is shown in figure \ref{f9} as red dashed line. The resulting absorption cross section is of the order of less than 1 mb, which is much smaller than what is usually obtained by different experiments. This apparent inconsistency will be discussed and resolved in section \ref{absorb}.

Determining a similar branching fraction of $N^* \rightarrow \Xi + K + K$ is not as straight forward as it is with the $\phi$ decay. First one should note that the pole masses of all $N^*$ resonances in UrQMD are below the threshold for this decay channel (2.3 GeV) and therefore the branching fraction will only be non-zero in the high mass tails of the resonances. Secondly there exists no experimental data on $\Xi$ production in elementary collisions near its production threshold. Therefore we use the new HADES data on $\Xi$ production in p+Nb reactions as a proxy for the unavailable elementary collision data to fix the $N^* \rightarrow \Xi + K + K$ branching fraction, as it is the dataset closest to an elementary reaction. To describe the measured production yield of $\Xi^-$'s (see table \ref{t1}) we obtain a branching fraction  $\frac{\Gamma_{\Xi+K+K}}{\Gamma_{\mathrm{tot}}} = 10 \%$ for all $N^*$ states mentioned above, that have a sufficiently high mass for this decay. A branching fraction of $10\%$ appears to be large, however one should keep in mind that this branching fraction applies only in the high mass tails of the resonances and the integrated fraction is less than one percent. Furthermore one expects that even higher mass resonances, not implemented in the model, do contribute stronger to the $\Xi$ production and one can think of this large branching fraction as the effective summed-up contribution of these high mass resonances. Table \ref{t1} summarizes the results for $\Xi^-$ production in $p+Nb$ at $E_{\mathrm{lab}}= 3.5$ GeV, measured with the HADES experiment as well as the results from our simulations.

\begin{table}[t]
\centering
\begin{tabular}{|c|c|}
  \hline
   \multicolumn{2}{|c|}{ HADES data } \\ \hline \hline	 
 	$\left\langle \Xi^- \right\rangle$ &	 $\Xi^-/\Lambda$  \\ \hline
   	 $(2.0 \pm 0.3 \pm 0.4)  \times 10^{-4}$  &  $(1.2 \pm 0.3 \pm 0.4) \times 10^{-2} $  \\ \hline \hline
   	   \multicolumn{2}{|c|}{UrQMD} \\ \hline\hline
  $\left\langle \Xi^- \right\rangle$  &  $\Xi^-/\Lambda$  \\ \hline 	  
$(1.44 \pm 0.05) \times 10^{-4}$  &  $(0.71 \pm 0.03) \times 10^{-2}$  \\ \hline    	  
		\end{tabular}
		\caption{$\Xi^-$ production yield and $\Xi^-/\Lambda$ ratio for minimum bias $p+Nb$ collision at a beam energy of $E_{\mathrm{lab}}= 3.5$ GeV, compared with recent HADES results \cite{Agakishiev:2015xsa}  }\label{t1}
\end{table}

\section{Sub-threshold $\phi$ and $\Xi$ production in nuclear collisions}

\begin{figure}[t]	
\includegraphics[width=0.5\textwidth]{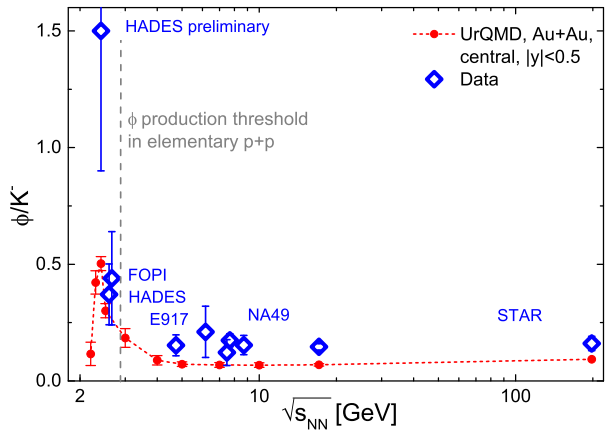}	
\caption{[Color online] Excitation function of the $\phi/K$ ratio for central ($b<5$ fm) Au+Au collisions, calculated with the UrQMD model including the new $N^*$ decays (red line). Experimental results from different beam energies and systems are shown as symbols \cite{Agakishiev:2009ar,Lorenz:2014eja,Piasecki:2014zms,Holzman:2001zh,Afanasev:2000uu,Adams:2004ux}. 
}\label{f3}
\end{figure}

Having constrained the branching fractions in the previous section, we employ this new mechanism to estimate the production probabilities of $\phi$'s and $\Xi$'s in nuclear collisions, particularly at sub-threshold energies. A ratio which has shown an interesting beam energy dependence, especially below the $\phi$ production threshold is the $\phi/K^-$ ratio, which is shown in figure \ref{f3} for nuclear collisions at different beam energies, measured by several experiments \cite{Agakishiev:2009ar,Lorenz:2014eja,Piasecki:2014zms,Holzman:2001zh,Afanasev:2000uu,Adams:2004ux}. Results from our simulations for most central ($b<3.4$ fm) Au+Au collisions are shown as the red line. Because the $\phi$ is an unstable particle it will never be directly measured in an experiment and therefore it is important to clarify how we define a measurable $\phi$ in our model simulation. We define a measurable $\phi$, for all the following results, as a $\phi$ which has decayed into a Kaon-Anti-kaon pair and whose decay partners have not rescattered. With rescattering we strictly mean no elastic or inelastic scattering of the decay products, noting that there can be a small fraction of $\phi$'s which can be reconstructed even though their decay daughter had an elastic scattering. However, we expect this correction to be on the order of a few percent.

From the comparison in figure \ref{f3}, it is clearly visible that the qualitative behavior of the data, a rapid increase of the $\phi/K^-$ ratio for sub-threshold energies, is nicely reproduced in our simulations. Also the value of the ratio is in nice agreement for beam energies at and above the HADES Ar+KCl data with $E_{\mathrm{lab}}= 1.76$ A GeV as well as the FOPI results for Ni+Ni collisions at $1.91$ A GeV. However, one also observes that above the low SPS energy regime the present model underpredicts the $\phi/K^-$ ratio. This can be understood as a result of the above mentioned high threshold for $\phi$ production in the string break-up. Because string excitation dominate the particle production at beam energies above $\sqrt{s_{NN}}> 5$ GeV, the $\phi$ must always be produced together with a Kaon-Antikaon pair, which strongly suppresses the $\phi$ production.\\ 
We also seem to underpredict the measured $\phi/K^-$ ratio for the lowest available beam energy of $E_{\mathrm{lab}}= 1.23$ A GeV, which is however still a preliminary result from the HADES collaboration \cite{Lorenz:2014eja}. An interesting feature of the calculations is the peak in the ratio at the aforementioned beam energy. The experimental confirmation of this peak, by $\phi$ measurements at even lower beam energies, would further support our approach for $\phi$ production in nuclear collisions.\\

At this point we have to note that the non-$\phi$ $K^-$ production in the UrQMD model, at the energies under investigation,
occurs via strangeness exchange reactions of the type $Y+\pi -> N+K^-$.
The cross sections of this process where taken from experimental data (see \cite{Graef:2014mra}).
Of course if the non-$\phi$ $K^-$ production is changed. e.g. due to a modification of the
exchange reaction cross section, then also the $\phi$/$K^-$ ratio will change, 
i.e. the $K^-$/$\pi^-$ and $K^-$/$K^+$ ratios, which should also change if only the
$K^-$ inelastic cross section is modified. These ratios will be discussed in the next paragraph.
An even more stringent test of the new $\phi$ production process is the measurement 
of the $\phi$ production cross section or total multiplicity (and not the ratio) in
the presented heavy ion experiments, which has been published for the Ar+KCl collisions by the HADES collaboration 
to be $\left\langle  \phi \right\rangle = (2.6 \pm 0.8) \cdot 10^{-4}$ \cite{Agakishiev:2009ar}. We obtain a total 
$\phi$ yield of $2.7  \cdot 10^{-4}$ which is in very good agreement with the HADES data. 

As already mentioned the ratio of $\phi/K^-$ not only depends on the production 
probabilities of the mesons but also on the absorption probability in the dense matter.
The inelastic cross section for $K^- + p$ (absorption) implemented in UrQMD is shown in figure \ref{f9} as black solid line. 
It follows from the measured (a fit to HERA data \cite{Graef:2014mra}) strangeness exchange processes and 
resonance excitations. 
We find that the UrQMD $K^-$ absorption cross section on protons is mainly consistent with the available data,
while at very low relative momenta the inelastic $K^- + p$ cross section is smaller than what has been measured. 
Therefore one could expect a to large $K^-$ yield at small beam energies.
The total production yield of $\phi$ and $K^-$ at the lower beam energy,
which would help to understand the different contributions to the peak in the ratio at
$E_{\mathrm{lab}}=1.23$ A GeV, 
has unfortunately not been published yet. 
Once this data is available we will be able to understand whether the difference in the observed ratio
is due to a to small $\phi$ or to large $K^-$ production yield.

\begin{figure}[t]	
\includegraphics[width=0.5\textwidth]{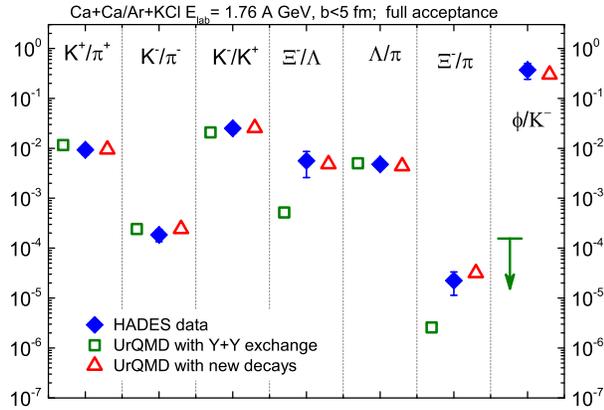}	
\caption{[Color online] Different strange particle ratios from the UrQMD model in its default settings (green squares), compared to our results including the new $N^*$ decays (red triangles). We compare the simulations for Ca+Ca at $E_{\mathrm{lab}}= 1.76$ A GeV and $b<5$ fm with published HADES data \cite{Agakishiev:2010rs,Tlusty:2009dk,Agakishiev:2009ar,Agakishiev:2009rr} for Ar+KCl collisions at the same beam energy (blue diamonds).  
}\label{f4}
\end{figure}

Finally we compare the multitude of strange particles produced in UrQMD, including the new $N^*$ decays, with sub-threshold nuclear collision data.  
In figure \ref{f4} we present results on strange particle ratios, in Ca+Ca collisions at $E_{\mathrm{lab}}= 1.76$ A GeV. The default calculation with the previously released UrQMD version (v3.4) is shown as as green squares. Compared to the default calculation we show the new results, including the $\phi$ and $\Xi$ decay channels of the $N^*$ as red triangles. A considerable increase in the $\phi$ and $\Xi^-$ production is visible. More importantly when we compare all the obtained strange particle ratios with Ar+KCl data from the HADES experiment (blue diamonds) we observe a very good description of all measured ratios, including the $\phi$ and $\Xi$. Such a good description of the full set of data has not been achieved in any previous study. Hence, we conclude that strange particle production in Ar+KCl collisions at the HADES experiment can be explained, and is in fully consistent, with production cross sections obtained in elementary reactions.   

\begin{figure}[t]	
\includegraphics[width=0.5\textwidth]{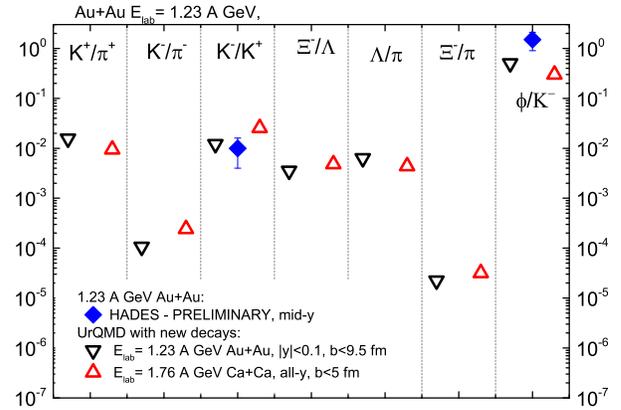}	
\caption{[Color online] The same particle ratios as in figure \ref{f4} for Au+Au reactions at $E_{\mathrm{lab}}= 1.23$ A GeV, calculated with the UrQMD model, including the new resonance decays. The red triangles correspond to the results for Ca+Ca, already shown in figure \ref{f4}. Preliminary HADES data \cite{Lorenz:2014eja} is shown as blue diamonds.
}\label{f5}
\end{figure}

In figure \ref{f5} we present predictions for the same strange particle ratios shown in \ref{f4}, in Au+Au reactions at 1.23 A GeV with the new $\phi$ and $\Xi$ production channels. The red triangles indicate the ratios for Ca+Ca collisions already shown in figure \ref{f4} for comparison, while the black triangles are the predictions for Au+Au collisions at an beam energy of $E_{\mathrm{lab}}= 1.23$ A GeV and $b<9.5$ fm. Collisions at this energy have been recently investigated at the HADES experiment. Up to now only preliminary data is available for few particle ratios, shown as blue diamonds. Apparently the preliminary $\phi/K^-$ ratio seems to indicate a small difference between data and our model study. It will be very interesting to see whether this holds for the final data and whether a similar difference will also be seen for the $\Xi^-/\Lambda$ ratio.

\section{Nuclear absorption of the $\phi$}\label{absorb}

An important aspect of $\phi$ production in nuclear collisions is the possibility of medium modifications of the
$\phi$ meson. A particularly interesting observable in this context is the so called nuclear transparency, which measures essentially the suppression (absorption) of $\phi$ meson production on nuclear targets of increasing mass number.
This observable is thought to be sensitive on the $\phi+N$ absorption cross section in a nuclear medium and should give
constraints on in-medium models of hadronic interactions.
First experiments measured the $\phi$ absorption in photo-production reactions at the CLAS and LEPS-SPring8 experiments \cite{Ishikawa:2004id,Wood:2010ei} to be of the order of 35 mb. A rather large $\phi+N$ absorption cross section of 14-21 mb was also inferred from hadronic production at the ANKE experiment \cite{Hartmann:2012ia}. These numbers appear to be much larger than the inelastic $\phi + p$ cross section which we obtained and is shown in figure \ref{f9}. However one should note
that these numbers where extracted using certain assumptions. For example the LEPS-SPring8 results used a Glauber model analysis and assumed that the $\phi$ is essentially produced instantaneously, while the ANKE analysis relied on comparisons to different models which may or may not be valid descriptions. It is therefore interesting to study the resulting $\phi$ suppression in nuclear targets within our model and compare it to available data.\\

\begin{figure}[t]	
\includegraphics[width=0.5\textwidth]{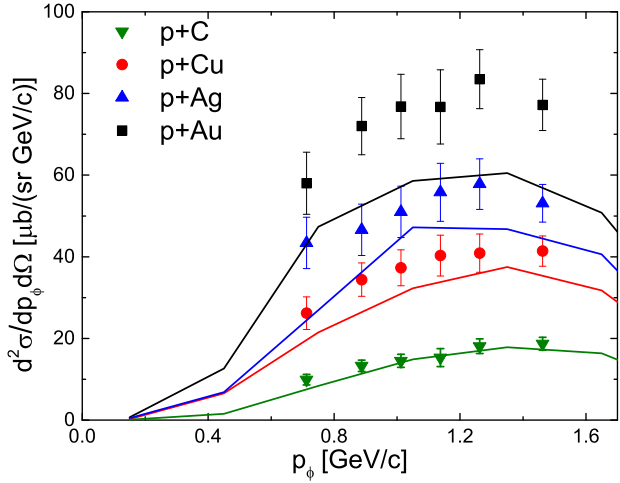}	
\caption{[Color online] The momentum dependent $\phi$ production cross section in different $p+A$ collisions at a beam energy of $E_{\mathrm{lab}}= 2.83$ GeV and within the ANKE acceptance (lines), compared to ANKE data from \cite{Hartmann:2012ia} (symbols).
}\label{f6}
\end{figure}

\begin{figure}[t]	
\includegraphics[width=0.5\textwidth]{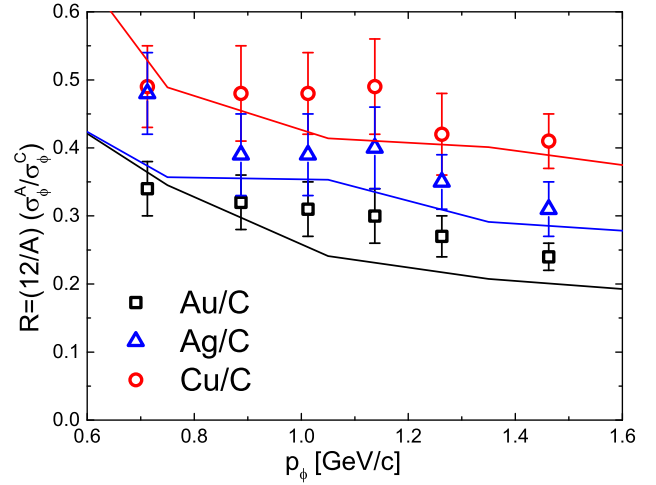}	
\caption{[Color online] The transparency ratio $R$ calculated within our approach (lines) compared to ANKE data from \cite{Hartmann:2012ia} (symbols). 
}\label{f7}
\end{figure}

Within our approach we cannot study photo-production of $\phi$ mesons, but we can investigate the proton induced reactions 
at the ANKE experiment.
Using the same parameters for the $\phi$ production by resonance as used in the previous section we 
calculated the momentum dependent $\phi$ production cross section in different $p+A$ collisions at a beam energy of $E_{\mathrm{lab}}= 2.83$ GeV and within the ANKE acceptance. The results are compared to the ANKE data in figure \ref{f6}.
We can see that the shape of the momentum dependent cross section is reproduced rather well, as well or even better than
other model results which include a $\phi$ meson modified in medium cross section \cite{Magas:2004eb,Paryev:2008ck,Hartmann:2010zzd}. The overall magnitude is underpredicted for the most heavy nuclear targets, while the $\phi$ production cross section on the lightest carbon nuclei is best described by our model.\\
The transparency ratio $R$:
\begin{equation}
R= \frac{12}{A} \frac{\sigma_{\phi}^{A}}{\sigma_{\phi}^{C}}
\end{equation} 
which is essentially the scaled ratio of the $\phi$ production cross section on different size nuclear targets is compared to ANKE
data in figure \ref{f7}. Again we observe that the ratio $R$ is well described by our model which does not include any explicit medium modification of the $\phi$. On the largest nuclear target, the cross section is even stronger suppressed in the model as compared to the data. The explanation of the strong suppression of the $\phi$ meson production in our approach cannot be a large inelastic $\phi+N$ cross section (which is of the order of $<1$ mb while the elastic cross section is set to 5 mb).
Because the $\phi$ is hardly absorbed (in our model) in the nuclear medium it is in fact the heavy mother resonance of the $\phi$ which rescatters with another nucleon before it can decay into $N+ \phi$. During this inelastic rescattering the resonance is likely to create one or two other resonances, however with smaller masses, which then are unable to decay into a $\phi$, effectively suppressing the $\phi$ production probability in a nuclear environment.
The same processes can be also responsible for the apparently large $\phi$ cross section measured in the LEPS-SPring8 experiment. If one assumes that also the photo produced $\phi$ mesons are also created through an intermediate resonance excitation, which can then easily rescatter and loose energy in the nuclear environment, we can also explain the apparent large $\phi+p$ cross section
in the LEPS-SPring8 experiment within our model.

\section{Conclusion}

In summary we have proposed and investigated a new mechanism of $\phi$ and $\Xi$ production in elementary and nuclear collisions, namely the decay of 
heavy resonances. For the $\phi$ production, the unknown branching ratios of the baryon resonances were extracted from $p+p\rightarrow p+p+\phi$ data measured by ANKE. The branching fraction necessary to describe the data is of the order of $0.2 \%$, in accordance with the measured violation of OZI suppression of $\phi$ production in p+p reactions.

Within our new approach to $\phi$ production we are able to describe the nuclear transparency ratio measured in proton induced reactions at the ANKE experiment, without the inclusion of any additional in-medium effects, and only a very small inelastic $\phi+p$ cross section. In our simulations the observed suppression of the $\phi$ production is due to the inelastic rescattering of the heavy mother resonance before it can decay into the $\phi$. This effectively suppressed the $\phi$ production in a nuclear environment. The large $\phi+p$ cross sections, extracted from the LEPS-SPring8 and ANKE experiments, might therefore be just a artifact of not taking into account the $\phi$ production processes.

For the $\Xi^-$ production the branching ratios of the heavy baryon resonances were extracted from p+Nb data of the HADES collaboration. Here a larger branching fraction of $10\%$ for $R^* \rightarrow \Xi + K + K$ is required. With this input from elementary reactions we are able to predict $\phi$ and $\Xi^-$ production in nuclear collisions at and below the elementary threshold energies. A good description of the HADES Ar+KCl data is achieved.

Even though a large branching fraction for the $N^* \rightarrow \Xi + K + K$ appears unlikely, one should keep in mind that the invariant mass required for the decay of $N^* \rightarrow \Xi +K +K$ is on the order of $2.3$ GeV, larger than the pole mass of any $N^*$ included in UrQMD. Therefore only resonances from the high mass tails may decay into a $\Xi$. Including resonances with pole masses larger than $2.3$ GeV in the model may therefore change the picture quantitatively.

In the future we intend to investigate the role of such higher mass resonances on particle production in elementary and nuclear collisions, with UrQMD. An important input for such a study would be measurements of $\Xi$ production rates in near threshold elementary collisions, giving direct constraints on heavy resonance production and decays. Consequently our study highlights the importance of resonance physics and dynamics in elementary and nuclear collisions in the energy regime of the SIS18 and the future SIS100 accelerator. Rare probes, like the multi-strange hadrons discussed in this paper can be very sensitive to unknown resonance states and their properties. Therefore if any conclusions on new physics are to be drawn from measuring such rare probes 
it is necessary to have a detailed understanding of the hadronic resonances and their dynamics in nuclear collisions.
 
\ack  
We would like to thank Tetyana Galatyuk and Christian Wendisch from the HADES collaboration for their help with the experimental data.
This work was supported by GSI and the Hessian initiative for excellence (LOEWE) through the Helmholtz International Center for FAIR (HIC for FAIR). The computational resources were provided by the LOEWE Frankfurt Center for Scientific Computing (LOEWE-CSC).


			\end{document}